# Spray-pyrolysis deposited La$_{1-x}$Sr$_x$CoO$_3$ thin films for potential non-volatile memory applications

## Z. Viskadourakis,[1,*] C. N. Mihailescu,[2] G. Kenanakis[1]

1. Institute of Electronic Structure and Laser, Foundation for Research & Technology-Hellas, N. Plastira 100, 70013, Heraklion, Crete, Greece.
2. National Institute for Laser, Plasma and Radiation Physics, 409 Atomistilor Street, P.O. Box MG-36, 077125 Magurele, Romania

*Corresponding author's e-mail: zach@iesl.forth.gr

## Abstract

In this work thin films of the La$_{1-x}$Sr$_x$CoO$_3$ (0.05 < x < 0.26) compound were grown, employing the so-called spray pyrolysis process. The as-grown thin films exhibit polycrystalline microstructure, with uniform grain size distribution, and observable porosity. Regarding their electrical transport properties, the produced thin films show semiconducting-like behavior, regardless the Sr doping level, which is most likely due to both the oxygen deficiencies and the grainy nature of the films. Furthermore, room temperature current-voltage (I-V) measurements reveal stable resistance switching behavior, which is well explained in terms of space-charge limited conduction mechanism. The presented experimental results provide essential evidence regarding the engagement of low cost, industrial-scale methods of growing perovskite transition metal oxide thin films, for potential applications in random access memory devices.

## Introduction

Among other perovskite transition metal oxides, Sr doped LaCoO$_3$ (LSCO) has gained considerable research interest, since it has been proposed for applications in fuel cells [1,2], oxygen penetration membranes [3,4], in gas sensing [5], and as electrodes in ferroelectric memory devices [6,7]. Very recently resistance switching (RS) behavior has been reported for LSCO [8,9] suggesting its possible application, to resistive random access memory devices (RRAMs). RS is not a bulk electronic property of LSCO system. In general, RS effect occurs in metal/perovskite oxide interfaces, and it is associated with the voltage-controlled crystalline defects occurring at the interface between the oxide and the metallic contact [10,11]. Thus, by manipulating the crystalline defects in the interface, through voltage pulses, accurately controllable RS devices can be produced, applicable in non-volatile memories.

In most of the above mentioned applications LSCO has to be grown in thin films. To this point, excellent quality LSCO thin films can be effectively grown, using pulse laser deposition (PLD) [12–14], magnetron sputtering [15,16] and metal-organic chemical vapor deposition (MOCVD) [17]. Such deposition methods, require sophisticated, and expensive experimental set-ups as well as highly dedicated personnel to use them. Moreover, thin films must be grown in high vacuum conditions, while the area of the deposited films lays in the range of a few mm$^2$. Therefore, those methods are highly appreciable in order to produce thin films for basic research purposes. Nonetheless, they are totally insufficient for large scale production, due to their exceptionally high cost, along with their low productivity rates. Towards the reduction of the production cost, preparation of the LSCO thin films using a simple process is highly desirable. To this point, several alternative methods have been developed. Among them, spray pyrolysis method is chosen as a simple, cost effective technique, which is used in both research laboratories, and industry to produce thin, uniform, smooth, crack-free, large area coatings, at the top of a great variety of substrates. In a brief description of this method, spray pyrolysis is based on forming a cloud of droplets (aerosol or spray) from a solution precursor, consisting of the appropriate metal salts. The produced aerosol is led at the top of a desired substrate, which is heated at an appropriate temperature. Thus the droplets are dried, while the heating of the precipitate favors the formation of particles of the desired composition.

Considering the above discussion, the growth of LSCO devices, appropriate for RS switching, using low-cost and user friendly techniques, becomes an interesting idea. In this context, we hereby present a thorough investigation regarding the thin film deposition of the LSCO compound (0.05 < x < 0.26), employing the well-known spray pyrolysis method. Using appropriate precursor solutions (described in the experimental part) thin films were deposited on Corning glass substrates and they have been characterized regarding their structural, crystalline and elemental properties. Furthermore, they have been subjected to electrical transport experiments. In particular, resistance was measured as a function of temperature and isothermal current vs. voltage experiments were being performed. From those experiments, RS behavior is evidently exhibited in Ag/LSCO interface, with notable endurance and retention characteristics. Further experimental data analysis, shows that the conduction is well interpreted in terms of space-charge-limited conduction (SCLC) mechanism. The presence of oxygen deficiencies along with grain boundaries both contribute to the existence of RS behavior. The presented results point towards the effective engagement of the spray pyrolysis route, for the fabrication of high-efficiency, non-volatile memory devices, based on transition metal oxides.

**Experimental details**

LSCO thin films (0.05 < x <0.26, nominal values) are deposited in Corning glass substrates, employing the so-called spray pyrolysis technique. In particular, highly pure La(NO$_3$)$_3$, Co(NO$_3$)$_2$ and Sr(NO$_3$)$_2$ (provided by Sigma-Aldrich) were used as starting materials. Appropriate amounts of them were dissolved in distilled water, under stirring of 1h. Upon stirring the solution was slightly heated. After stirring, the final solution (0.2 M) was left to cool down at room temperature. This solution is the precursor for the thin film deposition.

As mentioned above, flat Corning Eagle 2000 Borosilicate glass (Specialty Glass Products Inc.), with dimensions 10 mm × 4mm × 0.5 mm, were used as substrates. Before deposition, all substrates were ultrasonically cleaned in acetone (20 min), ethanol (20 min), and distilled water (20 min), and dried under flowing nitrogen, to remove any excess water traces. After that, they were immediately attached onto a horizontally positioned, hot-plate surface. The temperature of the hot-plate surface was controlled using a PID temperature controller.

Standard spray pyrolysis was employed, for the thin film deposition. More specifically, a small quantity of precursors, previously described, was introduced to a typical, commercially available air brush, which is commonly used for hobbies, crafts, fine art etc. The air brush was supplied by air pressure (~1 bar), through an oil-free, air compressor, while an electric valve was adapted between the air compressor and the air brush, in order to control the deposition time. The air brush nozzle (0.3 mm diameter) was ~10 cm far from the substrate, while the substrate temperature before starting the deposition was fixed at 350 °C. Upon the deposition, temperature decreased approximately by 10 °C, depending on the deposition time. In order to avoid such an effect we optimized the deposition technique as follows: Instead of depositing a thin film at once we performed short depositions along with long annealing time intervals, keeping the same total deposition time. For example, instead of continuously depositing for 1min, we deposit for 10 sec, stop for 1 min and repeat this sequence for 6 times. Air flow rate also affected the deposition temperature, however we kept it as low as possible by optimizing both the air brush pressure (~1 bar, as mentioned before) and the air brush nozzle diameter (diameter nozzle = 0.3 mm). Following this procedure, thin films of ~ 1000 nm are deposited. After deposition, all thin films were left on the hot-plate surface for at least 30 min and then they were cooled down to room temperature. After that, the samples were annealed at 700 °C for 2 h, in order the final LSCO phase to be formatted. . In the current study, we used both as prepared and fully oxygenated thin films. Oxygenation process takes place by heating the as-prepared films at 480 °C, for 2 h, under flowing oxygen.

Crystal structure of all produced samples was characterized by X-Ray diffraction (XRD) technique, using a Rigaku (RINT 2000) Cu *Kα* X-Ray diffractometer, while their surface morphology was studied by means of a field emission scanning electron microscope - SEM (FE-SEM, JEOL JSM-7000).

Raman measurements were performed at room temperature using a Horiba LabRAM HR Evolution confocal micro-spectrometer, in backscattering geometry (180°), equipped with an air-cooled solid state laser operating at 532 nm with 100 mW output power. The laser beam was focused on the samples using a 10× Olympus microscope objective (numerical aperture of 0.25), providing a ~14 mW power on each sample. Raman spectra over the 150-900 cm$^{-1}$ wavenumber range (with an exposure time of 10 s and 3 accumulations) were collected by a Peltier cooled CCD (1024 x 256 pixels) detector at -60 °C, with a resolution better than 1 cm$^{-1}$, achieved thanks to an 1800 grooves/mm grating and an 800 mm focal length. Test measurements carried out using different optical configuration, exposure time, beam power and accumulations in order to obtain sufficiently informative spectra using a confocal hole of 100μm, but ensuring to avoid alteration of the sample, while the high spatial resolution allowed us to carefully verify the sample homogeneity. The wavelength scale was calibrated using a Silicon standard (520.7 cm$^{-1}$) and the acquired spectra were compared with scientific published data and reference databases, such as Horiba LabSpec 6 (Horiba).

Further information regarding the chemical state of the deposited LSCO thin films were extracted from X-Ray Photoelectron Spectroscopy (XPS) experiments, which were performed using an ESCALAB Xi+ (Thermo SCIENTIFIC Surface Analysis) setup, equipped with a multichannel hemispherical electron Analyzer (dual X-ray source) working with Al K$_α$ radiation (hv= 1486.2 eV), using C 1s (284.4 eV) as the energy reference.

In addition, Ag contacts were deposited at the top of the thin films, by thermal evaporation technique. For contact optimization thin films were annealed at 400 °C, for 1 hour. Resistance measurements were performed in a home-built set-up in the temperature range 100 K - 300 K, employing conventional four-probe method. Furthermore, room temperature current vs. voltage measurements were done using another home-built microprobe station, with W tips (10 μm tip diameter), in combination with a Keithley 2400 SMU.

**Results and Discussion**

Fig. 1 presents the XRD spectrum of the x=0.26 thin film. Similar XRD patterns were obtained for all deposited thin films. XRD pattern shows diffraction peaks consistent with those of LaCoO$_3$ (JCPDS No.-01-084-0848). Among others, strong reflection is observed at $\sim 33°$ corresponding to the double peak 32.94° (110) and 33.30° (104) (Fig. 1b),characteristic for the $R\bar{3}c$ space group, in which the LaCoO₃ parent compound peaks are indexed in. No evidence of secondary phases can be observed, within the resolution of the measurement. Notably, the characteristic double peak at $2\vartheta \sim 33°$, exhibits a gradual transformation, towards a single peak (Fig. 1b), with increasing Sr. It has been previously mentioned that $R\bar{3}c$ is the space group of the parent compound LaCoO₃. In addition the fully oxygenated SrCoO₃, is crystallized to a cubic structure, with space group $Pm\bar{3}m$ [18]. Therefore LSCO exhibits a structural transition for $x \sim 0.55$ [19]. The Sr level of the deposited LSCO thin films, is far less than $x \sim 0.55$, and thus they are considered to have rhombohedral structure. Nevertheless, the gradual transition of the double peak at $2\vartheta \sim 33°$, towards a single one, is a bulk indication of the forthcoming structural transition.

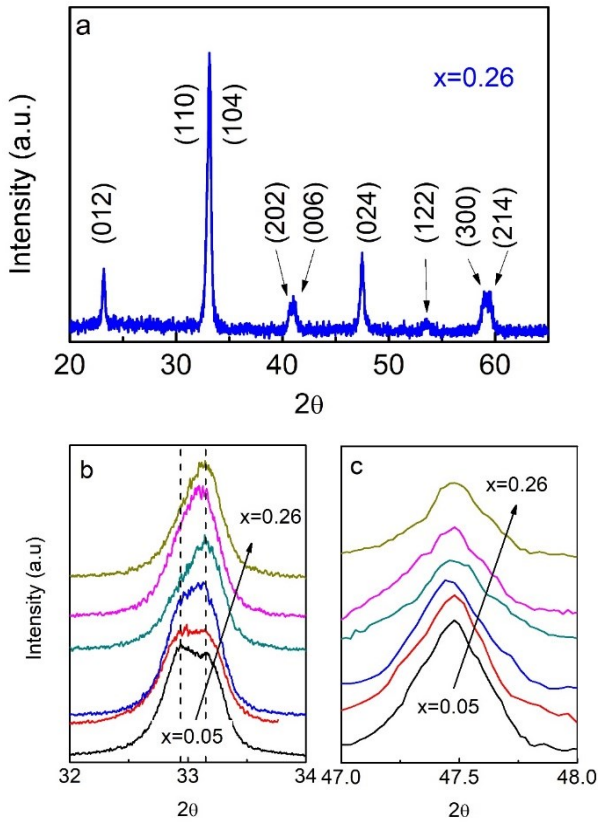

**Figure 1. a.** *XRD spectrum for LSCO thin film with x=0.26. Peaks have been identified in rhombohedral system. Similar spectra are obtained for all investigated thin films.* **b.** *Detailed examination of the double peak at ~33°. The double peak gradually transforms to a single, with increasing Sr content.* **c.** *peak at 47.5° which is used to estimate the crystallite size of all deposited thin films.*

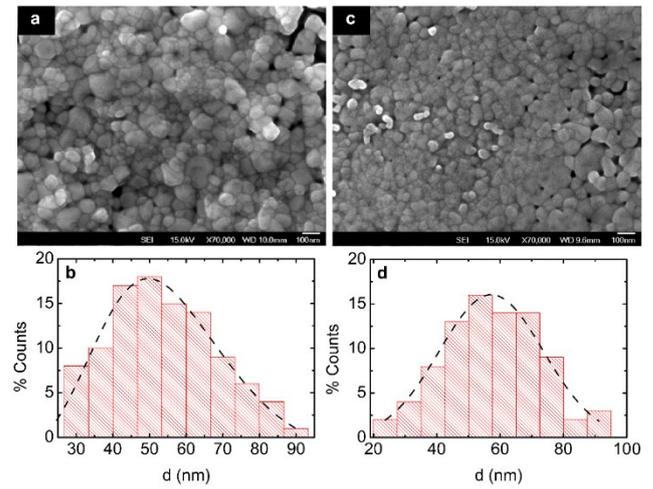

**Figure 2: a.** *SEM micrograph for thin film with x=0.07.* **b.** *Corresponding grain size distribution histogram* **c.** *SEM micrograph for x=0.19 thin film.* **d.** *Corresponding grain size distribution histogram. In both histograms, dash lines correspond to guides for the eye.*

The crystallite size was estimated using the Scherrer formula (D = 0.9 λ / β cosθ, where d, λ, θ, and β are the crystallite size, X-ray wavelength (1.5418 Å), Bragg diffraction angle, and full width of the half maximum (FWHM) of the diffraction peak, respectively. In our case, crystallite size of all samples was calculated from the broadening of the (024) reflection of the LSCO phase at 2θ = 47.5° (Fig. 1c) and it was found ~25 nm, irrespectively to the Sr doping. Such crystallite size is similar to those reported for polycrystalline powders sintered at 600 °C [20]. Therefore, the produced thin films are of single phase and they show polycrystalline characteristics.

Fig. 2a and c shows typical SEM micrographs of two investigated LSCO thin films, with x = 0.07 and x=0.19, respectively. Similar SEM pictures obtained for all deposited thin films. Both pictures show that thin films consist of grains. Corresponding statistical analysis (Fig. 2b and d), show that the average grain diameter is 55 – 57 nm. More details are presented in Table 1. The calculated grain size coincides with LSCO nanoparticles' grain size [21], which they have been synthesized employing complex chemical methods, and they have been subjected to sintering process, similar to ours. Furthermore, the grain size of our as-grown films is larger than that of LSCO film, deposited by flame spray pyrolysis [22].

| Sr doping | Mean grain size $d_{mean}$ (nm) | % grains with 45nm < $d_{mean}$ < 65 nm | % grains with $d > d_{mean}$ |
|---|---|---|---|
| 0.07 | 55 ± 13 | 50 | 67 |
| 0.19 | 57 ± 15 | 48 | 49 |

**Table 1:** *Statistical analysis results, extracted from histograms of figure 2 (panel b and d), for LSCO thin films, with x=0.07 and x=0.19 respectively*

Although, comparison between histograms and corresponding statistical analysis results reveal qualitative differences. More specifically, the grain size histogram of the x = 0.19 sample is fully symmetric, to the mean grain size value. On the contrary, the histogram of the x = 0.07 sample is asymmetric, i.e. ~ 67 % of the grains exhibit grain size larger than the average value. Generally, there are contradictory evidence regarding the effect of Sr doping to the LSCO grain size. For instance, for bulk LSCO samples, it has been reported that grain size increases with increasing Sr doping [23]. It has also been shown that processing conditions affect the grain size, causing changes in resistivity, thermal conductivity and Seebeck coefficient [20]. On the other hand, LSCO cathodes exhibit similar grain dimensions, regardless the Sr doping [24].

Although qualitative differences are observed between gran size distributions of x = 0.07 and x = 0.19 samples, there is not further evidence that they are attributed or even correlated to the Sr doping. Thus further, extensive investigation is needed towards this direction. Though, for the purposes of the current study, it is sufficient to conclude that all thin films exhibit similar grain sizes. Furthermore, empty space is observed between grains, thus thin films exhibit a sizable porosity, which is also consistent with previous reports [22]. Therefore, spray pyrolysis method is an effective method of developing porous, polycrystalline LSCO thin films, with nanometer-size grains.

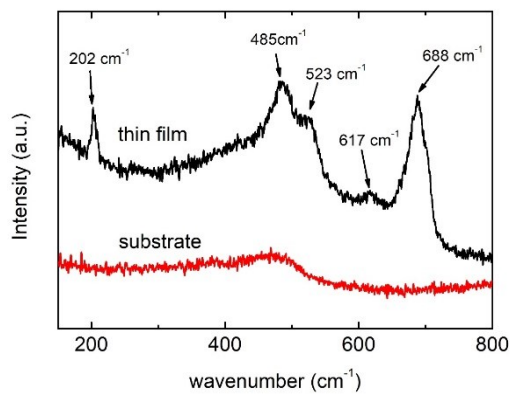

**Figure 3:** *Raman spectrum for x=0.12 thin film (black solid line). The background signal, coming from the substrate (red solid line) is also shown for direct comparison. Similar spectra are obtained for all deposited thin films.*

Figure 3 shows a typical Raman scattering response, for the investigated thin film x = 0.12. Among other features, several peaks are observed, at 202, 485, 522, 617 and 688 cm$^{-1}$. All of them can plausibly assigned to the perovskite LSCO structure. In particular, rhombohedral LaMnO$_3$ samples show peaks at 202, 490 and 611 cm$^{-1}$, which are quite close to our peaks at 202, 485 and 617 cm$^{-1}$, respectively [5]. Furthermore, the Raman peak at 522 cm$^{-1}$ can be assigned to the Eg quadrupole mode[25]. while the peak at 688 cm$^{-1}$ is assigned to the A$_{2g}$ breathing mode [26].

These two peaks are observed for epitaxial thin LSCO films, with thickness of 60-80 nm, while they are suppressed or even eliminated for either thinner or thicker films [27]. Interestingly the average nanoparticle size of our thin films nearly coincides with the thickness of those films. Even more, a broad hump in the range 400-440 cm$^{-1}$ is observed, which consists of multiple Raman peaks, coming from LaCoO$_3$ parent spin structure [25]. Similar Raman spectra are obtained for all investigated thin films, corroborating the highly pure, single LSCO phase of them.

Further information regarding the chemical state of the thin films, is obtained through XPS experiments. The XPS spectra of the Co 2p level, for all deposited thin films, is shown in Fig.4a. A broad peak is observed at ~780 eV, which is matched to the Co 2p$_{3/2}$ level, while another broad peak is shown at ~795 eV, which corresponds to the Co 2p$_{1/2}$ level, respectively. The energy difference, due to the spin-orbit splitting, is $\Delta E$ = 15.2 eV in all cases. Both binding energies and $\Delta E$ values are consistent to others reported in the literature [28–32]. Binding energies themselves cannot give sufficient information, regarding the Co oxidation state. However, such information can be extracted from the spin-orbit splitting energy difference $\Delta E$. In detail, it has been reported that $\Delta E$ = 16 eV for Co$^{2+}$ and $\Delta E$ = 15 eV for Co$^{3+}$, respectively [30, 32]. Thus the $\Delta E$ = 15.2 eV, obtained for Co 2p in LSCO thin films, can be attributed mainly to the Co$^{3+}$ [28, 29]. Moreover, another shoulder is observed at ~790 eV, which corresponds to low spin Co$^{3+}$ on the oxide surface [33]. Furthermore, the ~780 eV peak shifts to higher values with increasing Sr (Fig. 4b, main panel and upper, light yellow inset). Such an effect is also consistent to previously reported XPS results [28]. On the other hand, in the parent LaCoO$_3$ compound Co is stabilized as Co$^{3+}$. Sr$^{2+}$ substitution for La$^{3+}$ causes the formation of Co$^{4+}$, while a mobile hole is resulted for each Co$^{4+}$ ion. Such a Co oxidation state should be present in XPS spectra. The broad profile of both Co 2p$_{3/2}$ and Co 2p$_{1/2}$ definitely implies they consist of primary peaks. Indeed, a precise deconvolution of the Co 2p$_{3/2}$ peak (Fig. 4b, lower light blue inset) results to the appearance of other peaks, apart from the Co$^{3+}$ ones. For example, the peak at ~780 eV is analyzed to 4 consecutive sub-peaks, in the range 779.3 eV – 781.8 eV. The peak at 781.8 eV is satellite peak of the Co$^{3+}$ main one at 779.3 eV. The other two (at 779.8 eV and 781.8 eV) are defined as either due to the presence of Co$^{2+}$ or to Co$^{4+}$ ions. Considering these XPS experimental data, it is not possible to distinguish which of the two oxidation states produces those two peaks. However, Co$^{+2}$ could be discarded, since in that case a Co$^{2+}$ satellite peak should appear at binding energy ~785 eV [34 – 36]. Such a peak is not shown in neither of our XPS spectra. Considering that nominally Co$^{3+}$ and Co$^{4+}$ ions coexist in LSCO, Co$^{2+}$ ions cannot be present in LSCO compounds. Therefore, the peaks at 779.8 eV and 781.8 eV could possibly be due to the Co$^{4+}$ions. Though, further extensive investigation is needed to completely clarify such an issue.

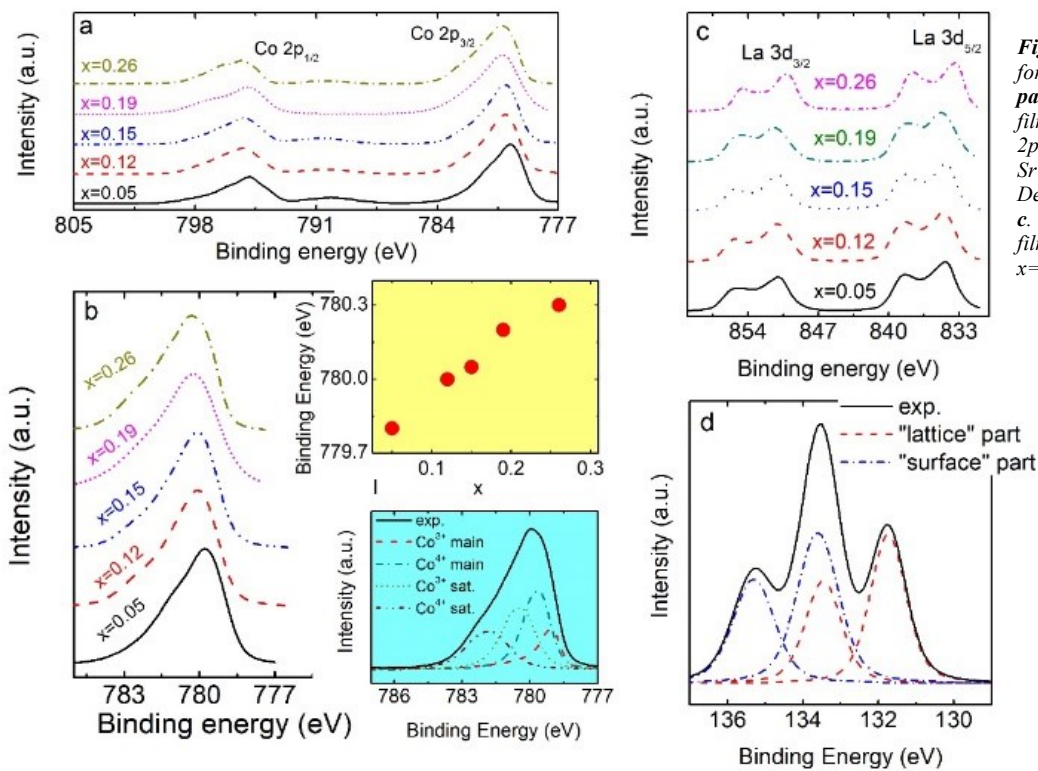

**Figure 4:** *Co 2p level XP spectra for all deposited thin films.* **b. main panel:** *Co 2p$_{3/2}$ peak for all thin films.* **upper, light yellow inset:** *Co 2p$_{3/2}$ binding energy as a function of Sr doping.* **lower, light blue inset:** *Deconvolution of the Co 2p$_{3/2}$ peak.* **c.** *La 3d XP spectra for all thin films.* **d.** *Sr 3d XP spectrum for the x=0.26 thin film.*

In addition the La 3d spectrum, for all thin films, are shown in Fig. 4c. Two main peaks are observed at 834.4 eV and 851 eV, corresponding to La 3d$_{3/2}$ and 3d$_{5/2}$ respectively. The energy difference between those peaks (originated from spin -orbit splitting) is 16.6 eV, consistently to the literature [32]. Furthermore, each peak is accompanied with another satellite peak, which is separated by 4 eV to the core peak, as reported to previous XPS studies [31, 32]. These satellite peaks are interpreted in terms of excitation of an electron from the anion valence band to the La f band [28, 32, 37]. Additionally, Sr doping does not result to any peak shift, in consistence to the literature [31].

Finally, the Sr 3d level XPS spectrum for the x = 0.26, is shown in Fig. 4d. The observed three peaks actually come from two separate 5/2 and 3/2 doublets, one with peaks at 136 eV and 134.2 eV (blue, dash dot line Fig. 4d) and another with peaks at 132.5 eV and 134.2 eV, respectively (red dash line, Fig. 4d). According to previous reports [38, 39], the former doublet can be assigned to either Sr ions, located at the surface termination layer of the thin film, or to secondary Sr-containing phases existed on the film surface and it is called "surface" component. The low energy doublet is also assigned to Sr ions, located just below the film surface, referred as "lattice" component. For both doublets the energy difference is ~2 eV, which is consistent to other reported values [40]. Similar spectra are also obtained for the other LSCO thin films.

Figure 5a, shows the resistivity $\rho$ as a function of temperature, for all investigated thin films. All thin films exhibit a semiconducting-like behavior, .i.e. the resistivity increases with decreasing the temperature. More specifically, all curves are well fitted to the equation $\rho(T) \sim \exp(T_o/T)^{1/4}$ [41], which means that Variable Range Hopping conduction dominates, in the temperature regime 140 K – 220 K (Fig. 5b), i.e. $T_o$ values, are presented in Table 2.

Essentially, in an insulator, charge carriers are localized in sites close to the fermi level, and each charge carrier can jump (hop) from one site to the other, when receiving enough energy, from an external field. According to the VRH model, the most probable hopping is not that between nearest sites. The most frequent hopping is that in which both the hopping energy and the hopping distance are minimized, simultaneously.

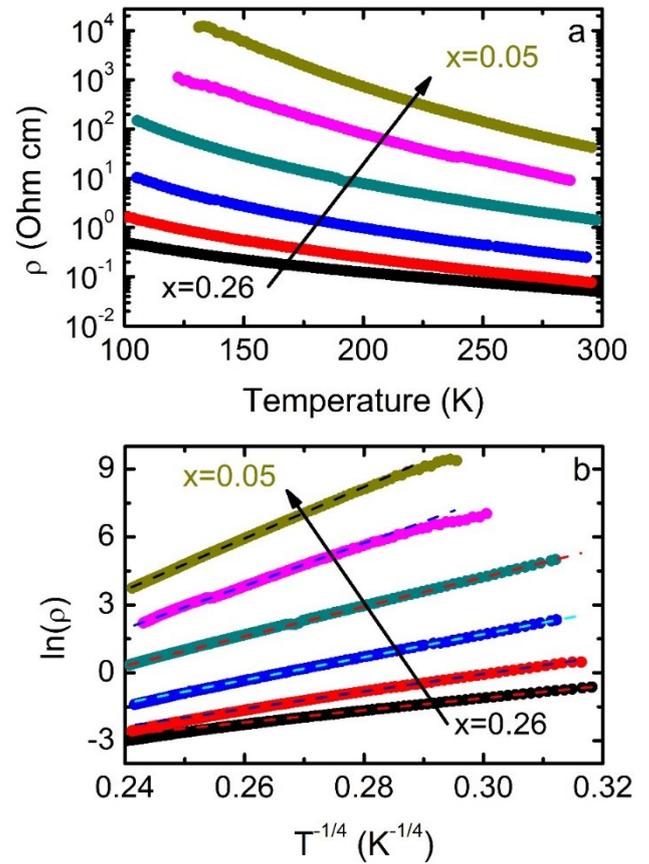

**Figure 5: a.** *Temperature dependence of the electrical resistivity for all deposited thin films.* **b.** *ln(ρ) vs. $T^{-1/4}$ curves for all films. Dash lines represent the corresponding VRH fittings.*

| x | T$_o$ (K) |
|------|----------------|
| 0.05 | 109.0 ± 0.3 |
| 0.07 | 93.1 ± 0.4 |
| 0.12 | 63.44 ± 0.05 |
| 0.15 | 51.51 ± 0.08 |
| 0.19 | 37.53 ± 0.04 |
| 0.26 | 27.23 ± 0.05 |

**Table 2:** *VRH fitting results for all deposited thin films.*

Furthermore, $\rho$ decreases with increasing Sr, indicating the introduction of charge carriers into the system by Sr doping, consistently to earlier reports, for single crystals, polycrystalline samples as well as for thin films [42 – 45]. As mentioned before, in the parent LaCoO$_3$ compound, Co cations are in 3+ oxidation state. Substituting Sr$^{2+}$ for La$^{3+}$ causes the formation of an equal number of Co$^{4+}$ cations. Increasing the oxidation state of Co ions induces holes in the system, resulting in a resistivity decrement. Nevertheless, at room temperature resistivity of the investigated thin films is rather high, comparing to previously reported values [45, 46]. Moreover it is well known that Sr doping results in the emergence of a metal-to-insulator transition (MIT) for x = 0.18 in the La$_{1-x}$Sr$_x$CoO$_3$ solid solution [42]. The Sr doping level of the samples investigated in this work lays in the range 0.05 < x < 0.26. Thus, samples with x < 0.18 are considered as that the enhanced resistivity values along with the semiconducting-like behavior of the LSCO samples with x > 0.18, might be attributed to both the oxygen deficiency and their well-defined grain nature.

The electrical transport properties of thin films are further investigated through isothermal current vs. voltage measurements. Typical I-V curve is shown in Fig. 6a, for x = 0.26 sample. At first, a polling voltage +10V is applied to the investigated device, which is used as a kind of device

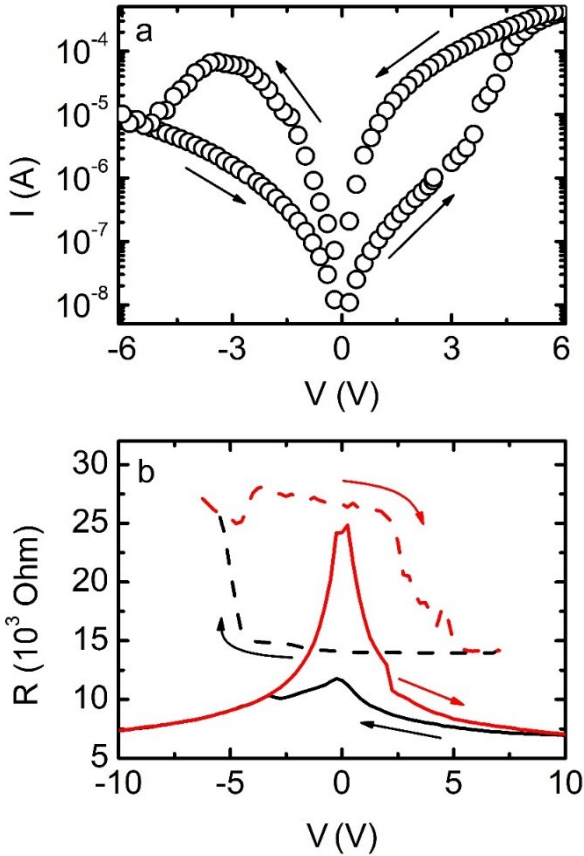

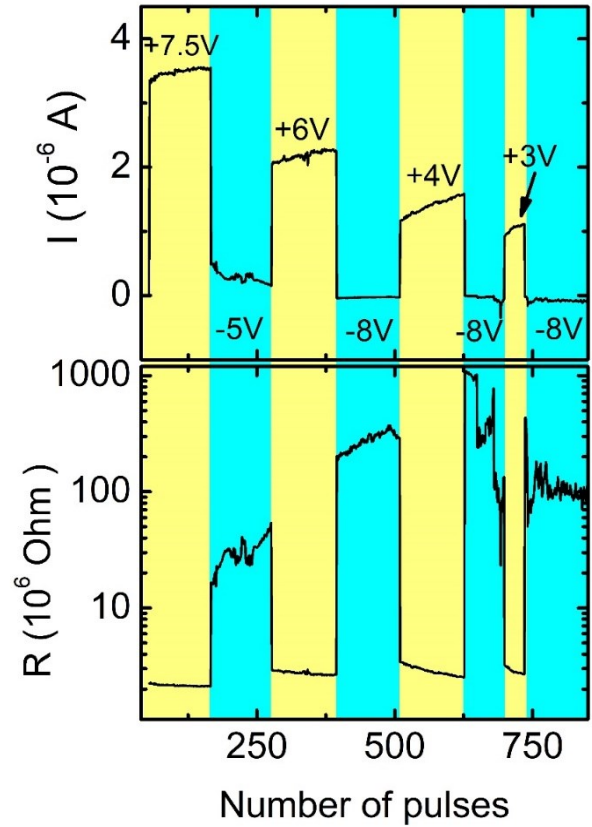

*Figure 6: a.* Current vs. Voltage for the x=0.26 thin film. Current is in absolute values. *b.* Resistance vs. Voltage curve for the x=0.26 thin film, according to ref. 8. Solid lines correspond to the resistance measured upon the application of "pump" voltage. Dash lines show the resistance obtained upon the application of the "probe" voltage.

*Figure 7: a.* Current vs. number of pulses for alternate voltages and *b.* corresponding resistance vs. time for the x=0.26 thin film. Current (resistance) measurements were taken every 0.5 sec.

insulating/semiconducting and therefore their resistivity could be semiconducting-like, however for samples with x > 0.18, the resistivity should increase with increasing temperature, indicating the metallic character of the material[27, 47, 48]. This contradictory behavior exhibited in the investigated thin films implies that apart from Sr doping, which affects the conduction, other conduction mechanisms are also highly involved. Oxygen vacancies, for example, could also contribute to the conduction of the LSCO [49, 50]. In particular each oxygen vacancy consumes two holes in LSCO and therefore the overall resistivity will be increased. Additionally, the grainy nature of all samples could contribute to conduction. In titanates [51], it has been observed that grain boundaries could result to the dramatic suppression of their conductivity. Furthermore, MIT as well as magnetic transition temperatures can be altered by varying the grain size, in manganites [52]. Therefore it could be plausibly assumed initialization. After that, the I-V was measured by sweeping the voltage in the sequence +V$_{max}$ → 0 → -V$_{max}$ → 0 → +V$_{max}$. The obtained I-V is characteristic of bipolar RS behavior. In particular, current sharply increases above V$_{SET}$ ~ 3.3 V, indicative for transition from high resistance (HR) to low resistance (LR) state, up to 6 V. Then, upon voltage decreasing the device persists in LR state, and for V$_{RES}$ ~ -4 V it returns back to the HR state. Furthermore, I-V curve shows asymmetry phenomena and hysteresis. In addition, the RS behavior is confirmed by applying a pump-probe procedure, previously described in the literature [8]. The results are presented in Fig. 6b, indicating two distinct resistance states, i.e. HR state in the negative voltage range and LR state in the positive voltage state, corroborating experimental evidence of fig 6a. In another experiment, positive and negative electric pulses (pulse duration ~200 msec) were alternatively applied to the device, and then the resistance was measured at 0.2 V, every 0.5 sec, for several times after each pulse. Corresponding results are presented in Fig. 7, indicating the well-defined difference between the HR and the LR states, although it is not stable, i.e. the R$_{HR}$

/ R$_{LR}$ ~ 20 - 300. Moreover, these results are indicative for good endurance and retention characteristics.

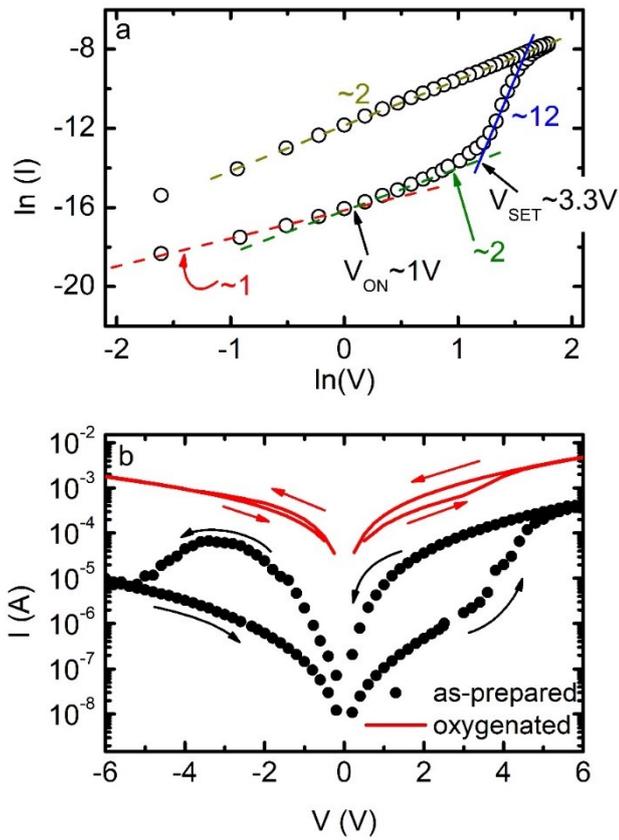

**Figure 8: a.** *ln(I) vs. ln(V) as extracted from Fig. 6a (positive voltage range). Dashed lines correspond to I~V$^n$ fittings, and arrows show both the V$_{ON}$ and V$_{SET}$. **b.** I-V curves for both as-prepared and fully oxygenated LSCO thin film with x=0.26.*

In general, such polarity dependent RS behavior commonly occurs in interfaces between metal electrodes and oxides and it is driven by electric fields [10]. Several models have been proposed to describe the conduction in electric-field induced RS devices, including Ohmic model, space-charge limited conduction, Poole-Frenkel emission etc. [53]. To distinguish among them, it is mandatory to look through the ln(I) vs. ln(V) curves, extracted from the corresponding I-V ones. In Fig. 8a, the ln(I) vs. ln(V) curve is presented, for positive voltages (as extracted from Fig. 6a). For low voltages (up to V$_{ON}$ ~ 1 V) the curve slope is ~1, thus Ohmic conduction occurs. That is, thermally generated free carriers dominate into the thin film. Then, slope changes to ~2, for voltages up to V$_{SET}$ ~ 3.3 V. Such slope implies that I ~ V$^2$, indicative for space-charge limited conduction (SCLC) [54]. In this voltage region excess charge carriers, which are injected from the contact to the thin film, are predominant over the thermally generated ones. Above V$_{SET}$, slope increases rapidly (~ 15), however it gradually decreases, and for voltages above ~ 5 V it is reduced back to ~ 2. Thus, injected carriers still dominate the conduction. Upon voltage reduction, slope remains ~ 2, thus SCLC still persists, and only for low voltages (~ 1 V) the slope approaches the ~ 1, thus the Ohmic behavior. Notably, no rapid current reduction is observed upon voltage reducing, which results to the hysteretic behavior in the RS. Such hysteresis indicates the existence of charge traps which affect the conduction [54]. Furthermore, in the voltage range 3.3 V – 5 V the slope large but continuously reducing, suggestive for exponential distribution of charge traps, regarding their energy level [55].

It is then obvious that SCLC model effectively describes the conduction in such LSCO/Ag interface, consistently to the literature [9]. More specifically, the SCLC mechanism is controlled by oxygen vacancies, which actually acts as charge traps, exponentially distributed in energy, absorbing injected carriers. In this context, we performed I-V experiments in fully oxygenated against non-oxygenated samples. Corresponding results are shown in Fig. 8b. For the oxygenated sample, hysteretic behavior has been suppressed. Oxygen vacancies, thus charge traps, have been reduced due to the oxygenation, resulting to the suppression of the hysteretic behavior. However, hysteresis is not eliminated, indicating the possible presence of other charge traps which contribute to the hysteretic RS behavior. In general, it has been proposed that any defect in the metal/oxide interface could potentially act as carrier trap, such as lattice defects (cation vacancies, dopants, interstitials, etc.) [56 – 58]. To this point, grain boundaries are imperfections, which could possibly contribute to the resistance switching behavior of the films. Such a grain boundary scenario has already investigated for several transition metal oxides [59 – 61]. Regarding the investigated LSCO thin films, interfaces between grains can be considered as energy barriers with height E$_b$, and width w. Moreover, the distance among neighbour grain boundaries can be modelled as the distance L between two energy barriers. All three parameters are directly affected by the grain size [62]. In each boundary an interface density of empty traps [63]. exists, which is assumed to be electrically neutral. As soon as first carriers are trapped, a potential barrier with height E$_b$, is created, which impedes the conduction of new-coming carriers with energy E < E$_b$. In other words, each barrier filters charge carriers with energy less than the barrier height. Although other probable scenarios cannot be excluded, the sustained hysteretic RS behavior of the fully oxygenated LSCO thin films, could possibly be interpreted as grain boundary effect.

Corresponding I-V curves have been obtained for all deposited LSCO thin films (Fig.9a-e), while appropriate experimental analysis, similar to that given for x = 0.26 sample, has been made. Apart from the x = 0.07 film, in which RS behavior cannot be observed (within the range of the applied electric field), all the other films show sizable RS behavior. Interestingly, films with x < 0.18 exhibit slight asymmetry and almost no hysteresis, in contrast to the films with x > 0.18, where both asymmetry and hysteresis are sizable. This experimental feature can also be attributed to the enhanced oxygen deficiency occurring in highly doped LSCO samples, in consistence to the literature [43]. In particular, in low Sr doped LSCO samples, the oxygen content is very close to the optimal value. By increasing the Sr doping (especially with x > 0.18), oxygen vacancies significantly increase as well, contributing to the observed RS behavior. Moreover, the evolution of the resistance switching, with respect to the Sr content reasonably confirms the grain boundary effect on the RS behavior, as previously proposed. Notably RS Furthermore V$_{RES}$ (the voltage, in which the LR to HR transition occurs) sharply changes from ~ -13 V to ~ - 5V as the x increases above the critical doping level x = 0.18, where the M-I transition takes place, suggesting that M-I transition directly affects the RS

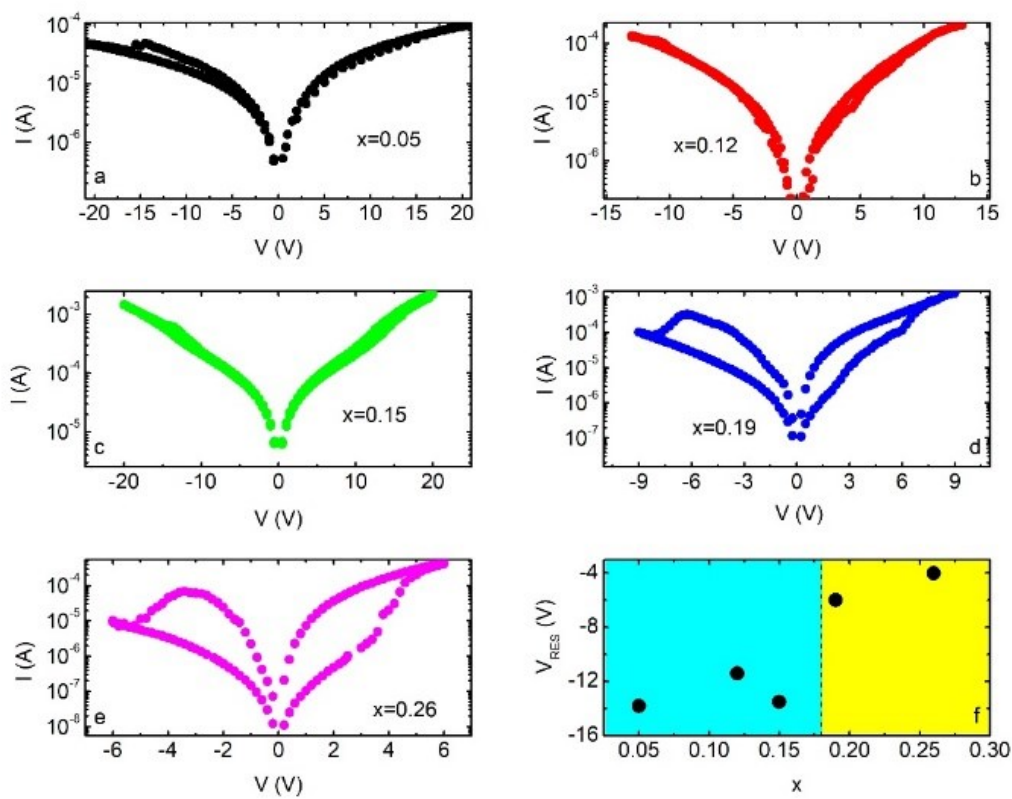

behavior of the device. Though, further investigation is needed to explore such an effect.

## Summary and conclusions

In the current study LSCO thin films were successfully grown, employing spray pyrolysis method. The produced thin films were characterized, regarding their structural, chemical and electrical transport properties.

XRD patterns of the LSCO thin films, reveal their single phase along with their polycrystalline character, while from the SEM pictures, the granular nature of the films was exposed. In particular, they exhibited a uniform grain size distribution, with average grain size of 55-57 nm, as well as a sizable porosity. Moreover, the obtained Raman spectra, show peaks corresponding to LaCoO₃ rhombohedral structure. Notably, similar peaks are observed for epitaxial LSCO thin films, with sample thickness comparable to the grain size of our thin films. On the other hand, chemical state of the deposited thin films was studied through, XPS experiments. Several peaks are observed, which all correspond to $Co^{3+, 4+}$ , $Sr^{2+}$ an $La^{3+}$ cations, in consistence with the nominal chemical formula of each thin film.

Resistivity vs. temperature was measured for all investigated thin films. All samples show a semiconducting-like behavior, regardless the Sr doping. Room temperature resistivity values are greater than the bulk ones. Moreover, for x = 0.18 bulk LSCO shows a M-I transition, however none of the films with x > 0.18 exhibit metallic behavior, indicating the possible contribution of both oxygen vacancies and grain boundaries to the conduction mechanism of the thin films. Nonetheless further investigation is needed to clarify which of the two dominates the system.

In addition, isothermal I-V curves were obtained for all investigated thin films. Almost all thin films exhibit resistance switch behavior, i.e. resistance possesses a low value for positive electric fields and a high value for negative electric fields. Such transport phenomena are common in metal/oxide interfaces. Conduction is effectively described through the space-charge limited conduction mechanism, controlled by charge traps. Oxygen vacancies mainly take the role of the charge traps in the current study, and upon their decreasing hysteresis observed in I-V curves, is suppressed, nevertheless it is not eliminated. The remained hysteresis in I-Vs could possibly be attributed to the grain boundaries which also can be considered as charge traps, which filter charge carriers, contributing to the conduction of the metal/oxide interface. Nevertheless comprehensive examination is required to deeply explore and distinguish between those two possible scenarios.

Conclusively, spray pyrolysis enables the construction of highly efficient LSCO resistance switch devices, engaging a user-friendly, low-cost and industry-applicable deposition method, pointing towards their massive production and use in novel resistive random access memories.

## Acknowledgements

This research has been co-financed by the European Union and Greek national funds through the Operational Program Competitiveness, Entrepreneurship and Innovation, under the call RESEARCH–CREATE–INNOVATE (project code: T1EDK-02784; acronym: POLYSHIELD).